\documentclass[twocolumn,apl,amsmath,amssymb,showpacs,superscriptaddress]{revtex4-2}
\usepackage{epsf} 
\usepackage{graphicx}
\usepackage{color}
\usepackage{soul}
\usepackage{gensymb}
\usepackage{sidecap}
\usepackage{amsmath}
\usepackage{mathtools}
\usepackage{float}
 \usepackage{multirow}
\usepackage[hidelinks,colorlinks=true,linkcolor=blue,citecolor=blue]{hyperref}
\DeclareUnicodeCharacter{2212}{\textendash}

\begin{document}
\title{Antiferromagnetic ordering and glassy nature in NASICON type NaFe$_2$PO$_4$(SO$_4$)$_2$} 
\author{Manish Kr. Singh}
\affiliation{Department of Physics, Indian Institute of Technology Delhi, Hauz Khas, New Delhi-110016, India}
\author{A. K. Bera}
\affiliation{Solid State Physics Division, Bhabha Atomic Research Centre, Mumbai 400085, India}
\affiliation{Homi Bhabha National Institute, Anushaktinagar, Mumbai 400094, India}
\author{Ajay Kumar}
\affiliation{Department of Physics, Indian Institute of Technology Delhi, Hauz Khas, New Delhi-110016, India}
\author{S. M. Yusuf}
\affiliation{Solid State Physics Division, Bhabha Atomic Research Centre, Mumbai 400085, India}
\affiliation{Homi Bhabha National Institute, Anushaktinagar, Mumbai 400094, India}
\author{R. S. Dhaka}
\email{rsdhaka@physics.iitd.ac.in}
\affiliation{Department of Physics, Indian Institute of Technology Delhi, Hauz Khas, New Delhi-110016, India}

\date{\today} 

\begin{abstract}
We investigate crystal structure and magnetic properties including spin relaxation and magnetocaloric effect in NASICON type NaFe$_2$PO$_4$(SO$_4$)$_2$ sample. The Rietveld refinement of x-ray and neutron diffraction patterns show a rhombohedral crystal structure with the R$\bar{3}$c space group. The core-level spectra confirm the desired oxidation state of constituent elements. The {\it dc}--magnetic susceptibility ($\chi$) behavior in zero field-cooled (ZFC) and field-cooled (FC) modes show the ordering temperature $\approx$50~K. Interestingly, the analysis of temperature dependent neutron diffraction patterns reveal an A-type antiferromagnetic (AFM) structure with the ordered moment of 3.8 $\mu_{B}$/Fe$^{3+}$ at 5~K, and a magnetostriction below $T_{\rm N}=$ 50~K. Further, the peak position in the {\it ac}--$\chi$ is found to be invariant with the excitation frequency supporting the notion of dominating AFM transition. Also, the unsaturated isothermal magnetization curve supports the AFM ordering of the moments; however, the observed coercivity suggests the presence of weak ferromagnetic (FM) correlations at 5~K. On the other hand, a clear bifurcation between ZFC and FC curves of {\it dc}--$\chi$ and the observed decrease in peak height of {\it ac}--$\chi$ with frequency suggest for the complex magnetic interactions. The spin relaxation behavior in thermo-remanent magnetization and aging measurements indicate the glassy states at 5~K. Moreover, the Arrott plots and magnetocaloric analysis reveal the AFM--FM interactions in the sample at lower temperatures. 
\end{abstract}

\maketitle
\section{\noindent ~Introduction}

In recent years, the search for novel sodium based oxide materials with exceptional magnetic and structural properties has been a driving force behind advancements in scientific and technological domains \cite{VSingh_PRB_23, RShukla_PRB_22, Saha_PRB_23}. Within this context, the NASICON (sodium superionic conductor) materials have emerged as captivating field of research, offering intriguing crystal structures and multifunctional properties \cite{Ouyang_NC_21}. Also, these materials have garnered significant attention due to their unique ability to conduct sodium ions, which paves the way for applications in energy storage, solid-state electrolytes, and electrochemical devices \cite{Simaran_WEE_21, Jian_AM_17}. The NASICON structure is characterized by a general formula of $A_x$$M_2$(XO$_4$)$_3$, where $A$ represents monovalent cations (e.g., Na, Li) with $x=$ 1--4, $M$ stands for transition metal cations which are octahedrally connected with ligands, and X represents tetrahedral network-forming anion (e.g., P, S) \cite{Ouyang_NC_21}. The distinctive crystal structure of these materials comprises 3D frameworks built from corner-sharing polyhedra. The $A$ and $M$ ions occupy interstitial sites, and the XO$_4$ groups form tetrahedral coordination environments for the $X$ atoms. The interstitial space created by this arrangement enable the migration of $A$ ions through two types of interstitial sites: $A$1 (one per formula unit) and $A$2 (three per formula unit), resulting in their excellent ionic conductivity \cite{Hagman_ACS_68}. These materials typically crystallize in rhombohedral ($R\bar{3}c$) \cite{Deng_CM_20} or monoclinic ($C2/c$) phase \cite{Zou_AFM_21} depending on the synthesis conditions and $A$ site occupancy. Moreover, their fascinating magnetic properties and structural versatility make them promising candidates for developing next-generation battery materials for cost effective energy storage  applications \cite{Chernova_JMC_11, LochabJSSC22}. 

Note that the magnetic properties of NASICON materials are intricately linked to the specific arrangement of the $M$ cations and their interactions with neighboring ligands. The coordination geometry of spin-active $M$ ions influences the magnetic behavior, including the presence of paramagnetic (PM), antiferromagnetic (AFM), or ferromagnetic (FM) ordering \cite{Jazouli_JPCS_88, Chernova_JMC_11}. Additionally, the structural flexibility of NASICON materials allows the incorporation of various magnetic dopants, thereby tailoring their magnetic properties and expanding their potential applications \cite{Idczak_JMMM_19}. The magnetic interaction between the $M$ cations present in neighboring octahedra is facilitated via super-super exchanges between the $M$ cations via O-X-O bonds as reported by Fanjat and Soubeyroux for Na$_3$Fe$_2$(PO$_4$)$_3$ using neutron diffraction study and found a long-range AFM ordering in monoclinic phase below 47~K \cite{Fanjat_JMMM_92}. The similar magnetic behavior of Fe$^{3+}$ was reported by Beltrán-Porter $et~al.$ using M$\ddot{\rm o}$ssbauer spectroscopy and magnetic susceptibility measurements, in which the antiferromagnetic transition occurred at 45.7~K with some weak ferromagnetism in Na$_3$Fe$_2$(PO$_4$)$_3$ sample \cite{Porter_RPA_80}. Moreover, Greaves $et~al.$ performed the  magnetization and neutron powder diffraction studies to demonstrate the magnetic behavior of Fe$^{3+}$ in Na$_3$Fe$_2$(PO$_4$)$_3$, but with a rhombohedral structure and found an anti-ferromagnetic transition below 47~K with a canted AFM ground state and a weak FM phase \cite{Greaves_PB_94}. 

Interestingly, the magnetic properties of Ti$^{3+}$ based NASICON material are explored within the scope of Figgis model and found that the Na$_3$Ti$_2$(PO$_4$)$_3$ exhibits a large deviation from the Curie-Weiss law below 100~K and undergoes a paramagnetic to antiferromagnetic transition below 50~K \cite{Jazouli_JPCS_88}. Within the same material class, the structural and magnetic behavior of Mn substituted Na$_3$Fe$_{2-x}$Mn$_{x}$(PO$_4$)$_3$ ($x=$ 0--0.4) was investigated through x-ray diffraction, dc and ac susceptibility, and Mössbauer spectroscopy \cite{Idczak_JMMM_19}. These samples crystallize in the monoclinic (C2/c) phase where both the Fe$^{3+}$ ($t_{2g}^3$$e_g^2$) and Mn$^{3+}$ ($t_{2g}^3$$e_g^1$) ions in the high spin state are located in octahedral coordination, and an antiferromagnetic transition was observed below 50~K. As the Mn content increases, the (Fe$_{1-x}$Mn$_{x}$)O$_6$ octahedra undergo distortion due to the Jahn–Teller effect of the Mn ions, which leads to another antiferromagnetic transition at T$_{\rm N2}$ = 43~K, with magnetic Fe ions having canted antiferromagnetic moments responsible for the transition at T$_{\rm N1}$=46.5 K \cite{Idczak_JMMM_19}. Also, the magnetic properties of alluaudite Na$_2$Fe$_2$(SO$_4$)$_3$, having monoclinic ($C2/c$) phase, exhibits long-range antiferromagnetic ordering below 12~K and the magnetic structure was explained by super-superexchange between Fe$^{2+}$ via SO$_4$ units using neutron diffraction analysis \cite{Dwibedi_API_16}. Moreover, the monoclinic phase of Na$_3$Fe$_3$(PO$_4$)$_4$ having Fe$^{3+}$ ions exhibit the long-range AFM ordering at $T_{\rm N}=$ 27~K \cite{Shinde_MRE_20}. The authors claim the absence of ferromagnetic interactions in the sample using the isothermal magnetization behavior and found no significant offset between FC and ZFC curves in magnetic susceptibility data \cite{Shinde_MRE_20}. Interestingly, in recent years the NASICON type mixed polyanionic NaFe$_2$PO$_4$(SO$_4$)$_2$ material, due to its structural tunability, is being considered a potential candidate as cathode for sodium-ion batteries \cite{Shiva_ENS_16, Yahia_JPS_18, Essehli_JPS_20}. However, the magnetic properties of NaFe$_2$PO$_4$(SO$_4$)$_2$ are vital to understand the complex magnetic interactions and have not been explored yet to the best of our knowledge. 

Therefore, in this paper we use the temperature dependent neutron diffraction (ND) and magnetization data analysis to probe the structural and magnetic ordering in NASICON structured polyanionic NaFe$_2$PO$_4$(SO$_4$)$_2$ sample. The crystal structure, vibrational and electronic properties are investigated using x-ray/neutron diffraction, Raman spectroscopy and x-ray photoemission spectroscopy, respectively. The detailed analysis of the ND patterns recorded in the temperature range of 5 to 300~K reveal an antiferromagnetic (AFM) ordering below 50~K, which found to be consistent with the {\it dc} and {\it ac} magnetic susceptibility behavior. We find a  magnetostriction below $T_{\rm N}=$ 50~K. The non-saturating trend in the M--H curve suggests the AFM ordering of the moments; however, the existence of low coercivity ($\sim$450~Oe) at 5~K suggests the presence of weak ferromagnetic (FM) correlations. The appearance of clear bifurcation in the magnetic susceptibility, measured between field-cooled (FC) and zero-field-cooled (ZFC) modes, indicates a glassy state at low temperatures. This found to be consistent with the behavior of aging and thermo-remanent magnetization (TRM) measured at 5~K. Also, the magnetic isotherm data are analyzed to examine the magnetic interactions and magnetocaloric properties of the sample. 

\section{\noindent ~Experimental}

The solution-assisted solid-state method is used to synthesize the NaFe$_2$PO$_4$(SO$_4$)$_2$ polycrystalline sample. The stochiometric amounts of NaNO$_3$ (0.605g, Merck, 99\%), Fe (NO$_3$)$_3$ 9H$_2$O (5.747g, Sigma-Aldrich, $>$98\%), and (NH$_4$)$_2$ SO$_4$ (1.8798g, Fisher Scientific, 99\%) were taken in 20 ml of DI water, and NH$_4$H$_2$PO$_4$ was taken in 10 ml DI water separately.  After the uniform mixing, the resulting solution was further mixed drop wise, and then heated at 80\degree C for 12 hrs to evaporate the water and got the yellow powder. The as-prepared yellow powder/pallet was annealed in the air for 12 hrs at 550\degree C. 

The neutron diffraction (ND) patterns are measured at different temperatures (5--300~K) at the powder diffractometer PD-1 ($\lambda$= 1.094~\AA) at Dhruva reactor, Trombay, India \cite{SKP_Parmana, SKP_NN}. For the ND measurements, the sample was filled in vanadium can of diameter 5 mm and attached with the cold head of a closed cycle helium refrigerator. The details of x-ray diffraction (XRD) measurements are given in \cite{supp_info}. The XRD and ND patterns are analyzed by the Rietveld refinement method using the Fullprof program package \cite{CARVAJAL_PBCM}. The Raman spectrum is measured at room temperature using the Renishaw inVia confocal microscope equipped with 532~nm, 2400 lines/mm grating, and 10mW power. The x-ray photoelectron spectroscopy (XPS) measurements are performed using the Kratos Analytical Ltd (AXIS SUPRA model) spectrometer with monichromatic Al K$_\alpha$  ($\lambda$= 1486.6 eV) source. The {\it dc} and {\it ac} magnetic susceptibility ($\chi-$T), the field dependent magnetization measurements (M--H), thermo-remanent magnetization (TRM), aging, and virgin magnetization curves are recorded using the MPMS-3 SQUID magnetometer from Quantum Design, USA. 

\section{\noindent ~Results and discussion}
 
 \begin{figure}[htbp]
\includegraphics[width=3.55in]{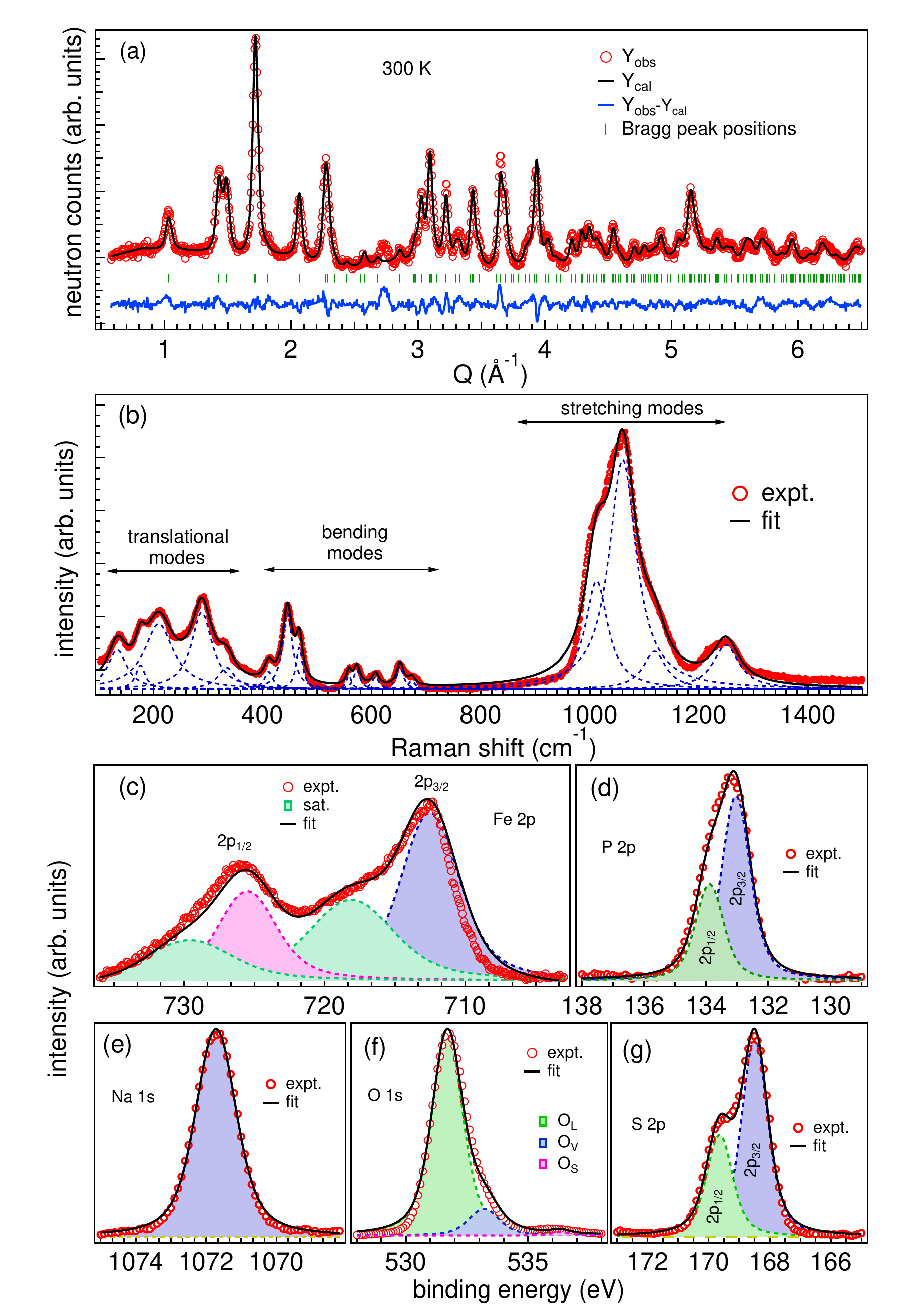}
\caption {(a) The experimentally observed (circles) and calculated (solid line) ND pattern for the NaFe$_2$PO$_4$(SO$_4$)$_2$ at 300~K. The solid line at the bottom represents the difference between observed and calculated patterns. The vertical bars indicate the positions of allowed nuclear Bragg peaks. (b) The Raman spectrum measured between 150--1500 cm$^{-1}$ and fitted with the Lorentzian profile. The room temperature core-level photoemission spectra of (c) Fe 2$p$, (d) P 2$p$, (e) Na 1$s$, (f) O 1$s$ and (g) S 2$p$.}
\label{F1_b}
\end{figure}
 
The Rietveld analysis of the ND pattern measured at 300~K reveals the crystal structure for NaFe$_2$PO$_4$(SO$_4$)$_2$ to be rhombohedral with space group R$\bar{3}$c, as shown in Fig.~\ref{F1_b}(a). The refined crystal structural parameters are given in Table I and the extracted lattice parameters $a=b=$ 8.4450(2) ${\rm \AA}$, $c=$ 22.0060(4) ${\rm \AA}$, $\alpha$= $\beta$=90\degree, $\gamma$=120\degree~and a unit volume of 1359.2(8) ${\rm \AA}^3$ are in good agreement with Refs.~\cite{Yahia_JPS_18, Shiva_ENS_16, Essehli_JPS_20}. Also, the Rietveld refinement of the XRD pattern is presented in Fig.~S1(a) of Ref.~\cite{supp_info}, which also confirms the formation of a single phase rhombohedral structure having R$\bar{3}$c space group. As neutrons are more sensitive for low atomic number elements in the structure as compared to the x-rays, the ND refinement indicates a slight deficiency of Na ($\sim$16\%) and oxygen ions ($\sim$9\%) in the sample. An unindexed peak at $\sim$ 2.7 \AA$^{-1}$ is observed, which may appear from a small secondary phase or sample environment. Further, we find that the Raman active modes observed between 900--1300 cm$^{-1}$ wave numbers, as shown in Fig.~\ref{F1_b}(b), are resulting from the intramolecular stretching (both symmetric and antisymmetric) of the PO$_4$ and SO$_4$ units  \cite{Ventruti_PCM_20, kloprogge_AM_02, Bih_JMS_2009, Bhalerao_PRB_12, Moreira_PRB_07}. More specifically, the peaks  around 1250 and 1060 cm$^{-1}$ can be assigned to the antisymmetric ($\nu_3$) and symmetric stretched vibrations ($\nu_1$) of SO$^{2-}_4$, respectively, \cite{Ventruti_PCM_20, kloprogge_AM_02}; whereas, those observed at $\sim$1110 and 1010 cm$^{-1}$ can be ascribed to the antisymmetric ($\nu_3$) and symmetric stretched ($\nu_1$) modes of PO$^{3-}_4$, respectively \cite{Bih_JMS_2009}. Further, several Raman active modes between 400--700 cm$^{-1}$ represent the symmetric and antisymmetric bending mode of PO$^{3-}_4$ and SO$^{2-}_4$ units. For example, the peak observed at $\approx$470 cm$^{-1}$ corresponds to the symmetric bending vibration ($\nu_2$) of SO$^{2-}_4$, while the peaks at $\sim$610 and 650 cm$^{-1}$ corresponds to the antisymmetric bending vibration ($\nu_4$) of SO$^{2-}_4$ \cite{Ventruti_PCM_20, kloprogge_AM_02}. Moreover, the asymmetric bending modes ($\nu_4$) corresponding to PO$^{3-}_4$ are obtained at 560 and 570 cm$^{-1}$, and symmetric bending mode ($\nu_2$) is observed at 445 cm$^{-1}$ \cite{Bih_JMS_2009}. Also, the Raman modes at the low wave number between 100--400 cm$^{-1}$ are ascribed to the translational/coupled-transitional motion of Fe$^{3+}$ ion and with its neighboring atoms \cite{karthik_JMS_20}. 

\begin {table}
\caption {The atomic positions and the site occupancies of NaFe$_2$PO$_4$(SO$_4$)$_2$ sample, as extracted from the Rietvield analysis of the neutron diffraction pattern at 300~K.} 
\label{tab:title} 
\begin{center}
\begin{tabular}{ p{1cm}p{1.2cm}p{1.4cm}p{1.3cm}p{1.3cm}p{1.3cm}}
 \hline
 \text{Atom} & \text{Wyckoff}&\textit{x}&\textit{y}&\textit{z}&\text{Occup.}\\
 \hline
 Na1  &     6b &    0.00000  &  0.00000 &   0.0000&    0.84(3)   \\
 Fe1   &   12c  &     0.00000 &   0.00000 &   0.1479(2) &   1.00   \\
 S1    &  18e  &    0.2938(11)  &  0.00000   & 0.2500  &  0.67    \\
 P1    &     18e &  0.2938(11) &   0.00000 &   0.2500 &   0.33    \\
O1     &    36f  &  0.0264(9)   & 0.2113(8) &   0.1946(2)  &  0.93(1)   \\
O2     &   36f  &  0.1912(8)   & 0.1686(8) &   0.0884(2)  &  0.89(1)   \\
 \hline
\end{tabular}
\end{center}
\end {table}

To understand the electronic properties of the NaFe$_2$PO$_4$(SO$_4$)$_2$ sample, in Figs.~\ref{F1_b}(c--g), we show the core-level photoemission spectra of constituent elements. The calibration of all the spectra is done with respect to the C 1$s$ peak at 284.6~eV. After subtracting the inelastic background using Tougaard method, the Na 1$s$, P 2$p$, Fe 3$d$, O 1$s$, and S 2$p$ core-levels are fitted using the Voigt function (in IGOR Pro software), which is the convolution of the Gaussian and Lorentzian profiles, with a shape factor of 0.6. The Fe 2$p$ spectrum, in Fig.~\ref{F1_b}(c), consists of 2$p_{3/2}$ and 2$p_{1/2}$ peaks along with their respective satellite peaks. These core-level peaks are observed at 712.5~eV and 725.5~eV with their corresponding satellite features at 718.1~eV and 729.6~eV, which indicate the presence of Fe solely in the trivalent state \cite{castro_JPCC_10, Yao_JAC_21}. In Fig.~\ref{F1_b}(d), the peaks centered at 133.0~eV and 133.9~eV are attributed to the spin-orbit splitting of P 2$p$, i.e., the 2$p_{3/2}$ and 2$p_{1/2}$ levels, respectively, which correspond to the +5 oxidation state of phosphorus in the tetrahedral environment \cite{castro_JPCC_10}. Fig.~\ref{F1_b}(e) displays the Na 1$s$ core level centered at 1071.1~eV, which indicate the monovalent state of sodium \cite{Citrin_PRB_73}. The deconvoluted O 1$s$ spectrum is depicted in Fig.~\ref{F1_b}(f), revealing the presence of three components located at (i) 531.7~eV (O$_L$), which can be assigned to the lattice O$^{2-}$, (ii) 533.2~eV (O$_V$) is attributed to the oxygen vacancies (estimated $\sim$8\% by taking area under the respective peaks), which are consistent with the neutron analysis, and (iii) 536.3 eV (O$_S$) corresponding to a surface hydroxyl group (-OH) \cite{Citrin_PRB_73, Ajay_JAP_20}. Fig.~\ref{F1_b}(g) exhibits two characteristic peaks at 168.5~eV and 169.6~eV corresponding to the S 2p$_{3/2}$ and S 2p$_{1/2}$ energy levels, respectively associated with the SO$_4^{2-}$, which are attributed to 5+ oxidation state of sulfur in the tetrahedral coordination \cite{Yao_JAC_21}. 

\begin{figure}[htbp]
\includegraphics[width=3.1in]{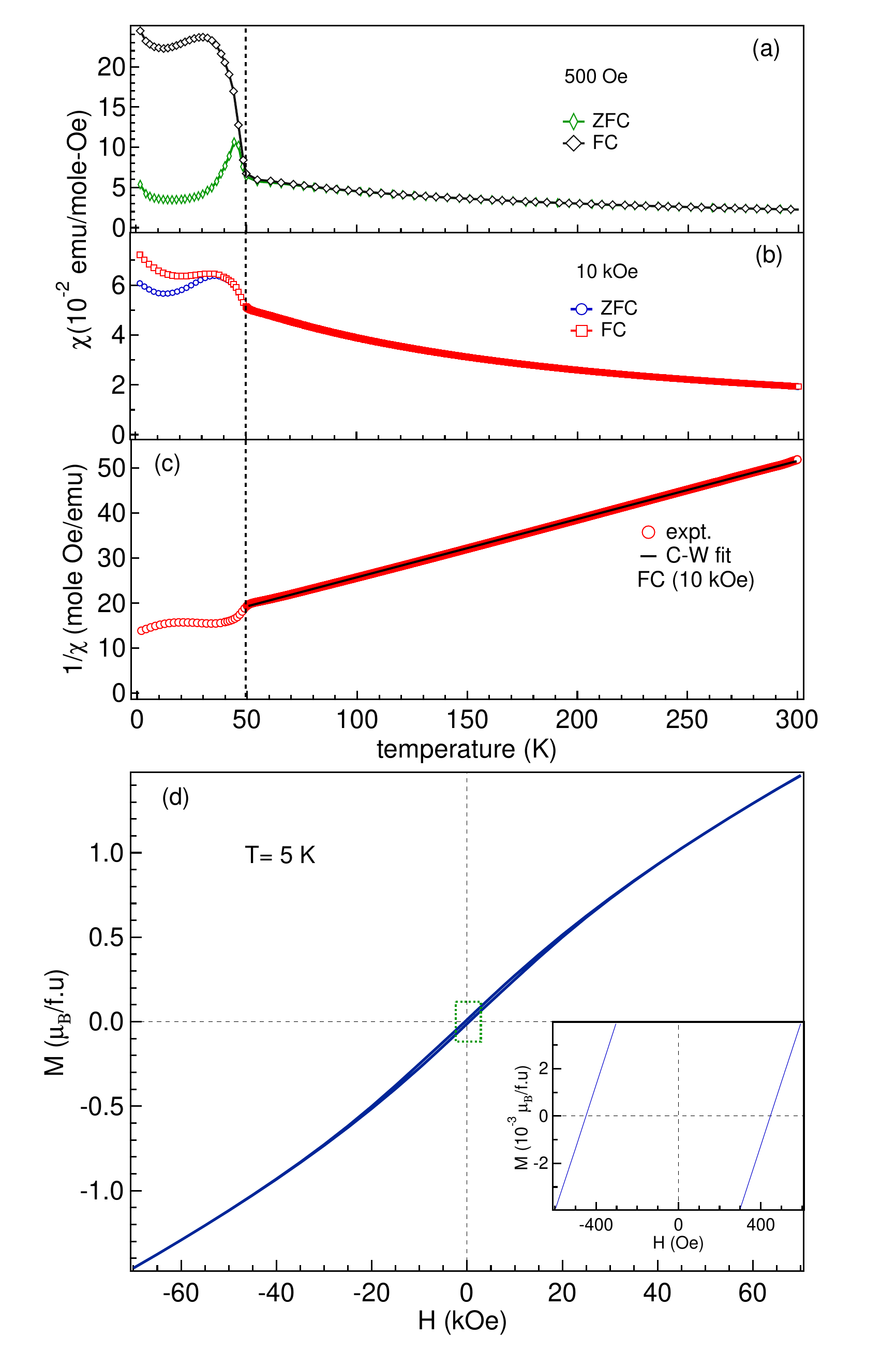}
\caption {The {\it dc} magnetic susceptibility of NaFe$_2$PO$_4$(SO$_4$)$_2$ sample in ZFC and FC modes from 2 to 300~K measured at (a) 500~Oe and (b) 10~kOe. (c) The inverse magnetic susceptibility of the FC curve where the black solid lines represent the Curie-Weiss fitting from 60--300~K. (d) The M--H curve in ZFC mode measured at 5~K where the inset shows the enlarged view in the low field region.}
\label{F2_R}
\end{figure}

In order to investigate the magnetic interactions in the NaFe$_2$PO$_4$(SO$_4$)$_2$ sample, we measure the temperature dependent {\it dc} magnetic susceptibility ($\chi= M/H$) from 2 to 300~K in both ZFC and FC modes at 500~Oe and 10~kOe, as depicted in the Figs.~\ref{F2_R}(a, b), respectively. We find an abrupt increase in the $\chi$ values below 50~K (indicated by the vertical dashed line). Interestingly, there is a large bifurcation between ZFC and FC curves below $\approx$45~K, with a significant suppression and shift in bifurcation at 10~kOe [see Figs.~\ref{F2_R}(a, b)] providing an evidence of the presence of glassy nature of the magnetic phase in this sample at low temperatures \cite{Sampathkumaran_PRB_2002, Ajay_PRB_20_1}. The FC curve reaches a maximum around 30~K and then show a downturn till $\approx$20~K, which are characteristics of the AFM ordering \cite{Saha_PRB_23}; however, a slight upsurge at very low temperature is probably due to paramagnetic impurities \cite{Saha_PRB_23, Guchhait_PRB_22, Yogi_PRB_17}. In Fig.~\ref{F2_R}(c), the inverse of magnetic susceptibility (measured in FC mode at 10~kOe field) is plotted, and the high-temperature range (60--300~K) is fitted using the Curie-Weiss (C-W) law $\chi=C/(T-T_\theta$) to find the effective magnetic moment ($\mu_{\rm eff}$), where C is the C-W constant and $T_\theta$ is the C--W temperature. The extracted values of  $T_\theta$ and the $\mu_{\rm eff}$ are found to be about -98~K and 3.93 $\mu_{B}$/Fe$^{3+}$. The value of $\mu_{\rm eff}$ is found to be close to the spin-only magnetic moment (3.87 $\mu_{\rm B}$) for the intermediate spin (IS) state of Fe$^{3+}$ ions ($t_{2g}^4$$e_g^1$; S=3/2), calculated using the formula $\mu_S = 2 \sqrt{S(S+1)}~\mu_B$, which suggests the existence of Fe$^{3+}$ ions fully in the IS state or in mixed of high ($t_{2g}^3$$e_g^2$; S=5/2) and low ($t_{2g}^5$$e_g^0$; S=1/2) spin states in 52 to 48 ratio \cite{Shankar_Nat_18, Toney_Inorg_84, Ming_JAP_12}. The negative value of $T_\theta$ obtained from the C--W fit suggests the existence of antiferromagnetic interactions in the system \cite{Guchhait_PRB_22}. Also, a collinear AFM ordering below $T_{\rm N}$= 10.4~K is reported in an isotropic triangular lattice of Na$_3$Fe(PO$_4$)$_2$ in monoclinic phase ($C2/c$) and having the high spin Fe$^{3+}$ ions \cite{Sebastian_PRB_22, Ambika_JPCM_23}. Moreover, the unsaturated behavior in the field-dependent magnetization, M-H loops, even up to $\pm$70 kOe indicates the AFM/uncompensated spins at 5~K, as shown in the Fig.~\ref{F2_R}(d). However, a symmetric M--H loop is observed with small coercivity (H$_{\rm C}$) of around 450~Oe, as displayed more clearly in the inset of Fig.~\ref{F2_R}(d), which suggests the presence of weak FM type interactions at 5~K. These observations motivated us to further investigate the magnetic correlations in the sample by detailed analysis of neutron diffraction and magnetizaton in different protocols at low temperatures.  

\begin{figure}[htbp]
\begin{center}
\includegraphics[width=3.45in]{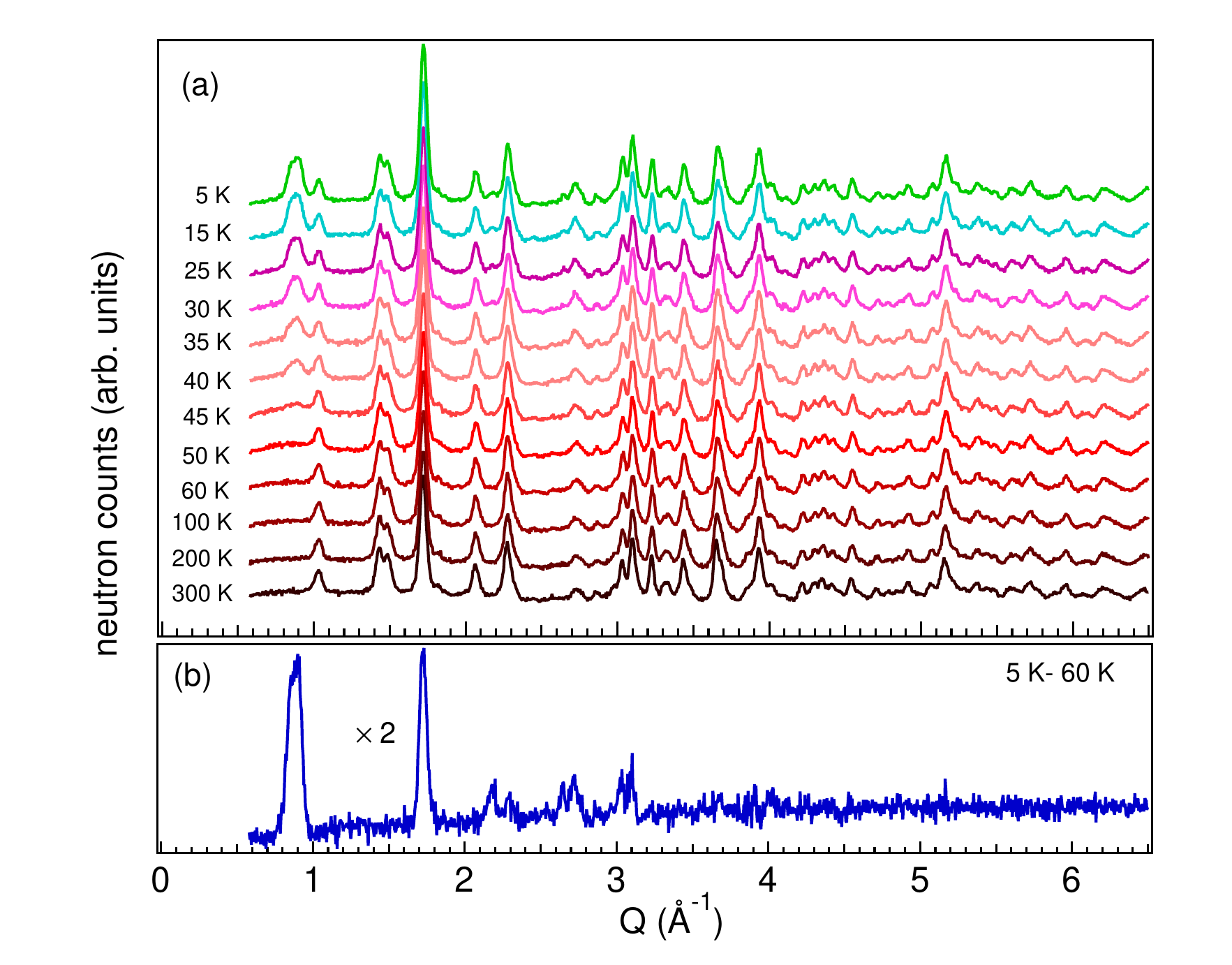}
\caption {(a) The ND patterns measured in the temperature range of 5--300~K. (b) The magnetic diffraction pattern at 5~K after subtraction of the nuclear background at 60~K.}
\label{F3}
\end{center}
\end{figure}

\begin{figure*}[htbp]
\includegraphics[width=7.35in]{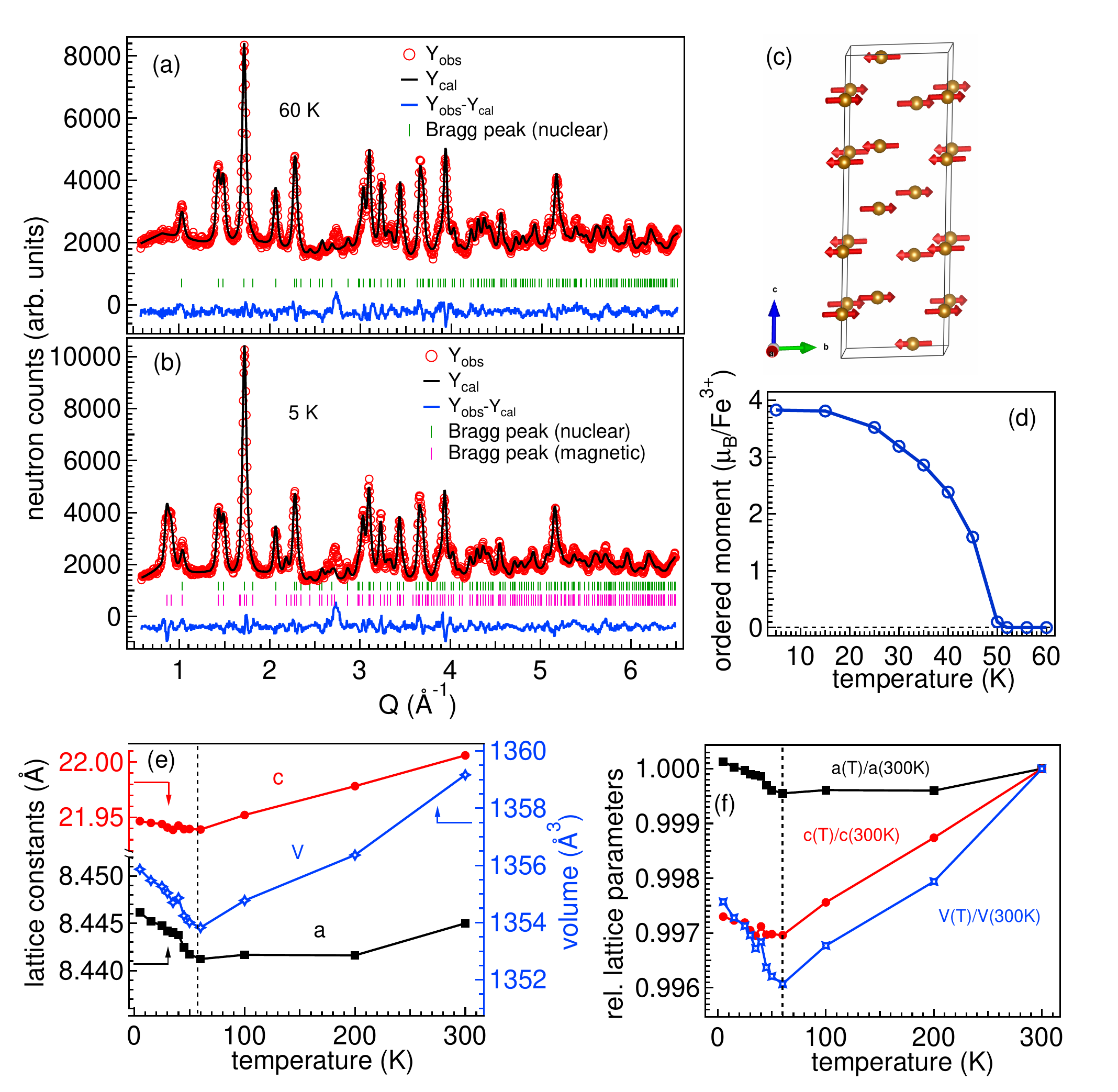}
\caption {The experimentally observed (circles) and calculated (solid line) ND patterns for the NaFe$_2$PO$_4$(SO$_4$)$_2$ at (a) 60~K and (b) 5~K. The solid lines at the bottom of each panel represent the difference between observed and calculated patterns. The vertical bars indicate the positions of allowed nuclear and magnetic Bragg peaks. (c) A schematic of the magnetic structure of NaFe$_2$PO$_4$(SO$_4$)$_2$ sample. (d) The ordered magnetic moment (order parameter) of Fe$^{3+}$ ion as a function of temperature. (e) The temperature dependent lattice parameters $a$ and $c$, and the unit cell volume $V$, and (f) the lattice parameters $a$ and $c$, and the unit cell volume $V$ normalized with their respective value at 300~K.}
\label{F4}
\end{figure*}

To further investigate the magnetic ordering at low temperatures, in Fig.~\ref{F3}(a) we present the as measured ND patterns in the large temperature range of 5--300~K. Interestingly, the nature of ND patterns remains the same with an appearance of additional magnetic Bragg peaks at $\sim$0.85, 0.9, and 2.16 \AA$^{-1}$ along with an increase of the intensity of the nuclear Bragg peaks at $\sim$1.72, 2.72, 3.0, and 3.1 \AA$^{-1}$ below the T$_{\rm N}=$ 50~K. The pure magnetic ND pattern at 5~K is estimated by subtracting the nuclear background measured at 60~K (PM state). The appearance of new magnetic Bragg peaks, as shown in Fig.~\ref{F3}(b), confirms an AFM ordering. 

Note that all the magnetic Bragg peaks could be indexed with a propagation vector $k=$ (0 0 0) with respect to the rhombohedral nuclear unit cell. The symmetry-allowed magnetic structure is determined by a representation analysis using the computer program SARAh-Representational Analysis \cite{WILLS_PB_2000}. The representational analysis \cite{Bertaut_JAP_1962, Bertaut_ACCP_1968, Bertaut_JPC_1971, Bertaut_JMMM_1981, Yu_springer_91, Bradley_Oxford_72, Bera_PRB_22, Bera_PRB_17, Bera_PRB_16, Dutta_PRB_22, Suresh_PRB_18} allows the determination of the symmetry-allowed magnetic structures that can result from a second-order magnetic phase transition, given the crystal structure (R$\bar{3}$c) before the transition and the propagation vector [$k=$ (0 0 0)] of the magnetic ordering. They involve first the determination of the space group symmetry elements, $g$, that leave the propagation vector $k$ invariant: these form the little group G$_k$. The space group R$\bar{3}$c involves three centering operations and twelve symmetry operations and out of them, twelve leave the propagation $k$ invariant or transform it into an equivalent vector. The symmetry analysis reveals that there are six irreducible representations (IR), i.e., six possible symmetry-allowed magnetic structures. Out of the six IRs, four IRs ($\Gamma_1 - \Gamma_4$) are 1D and the two IRs ($\Gamma_5 - \Gamma_6$) are 2D. All the six IRs are nonzero for the magnetic site (12$c$) of the present compound and therefore, the magnetic representation of a crystallographic site can then be decomposed in terms of the IRs as below: 
\begin{equation}
\Gamma_{M a g}=\sum_v n_v \Gamma_v^\mu=\Gamma_1^1+\Gamma_2^1+\Gamma_3^1+\Gamma_4^1+\Gamma_5^2+\Gamma_6^2
\end{equation}
The basis vectors of six IRs (the Fourier components of the magnetization) for the magnetic site are given in Ref.~\cite{supp_info}. The labeling of the propagation vector and the IRs follows the scheme used by Kovalev \cite{Kovalev_GBSP_93}. 

Interestingly, out of the above IRs, the best refinement of the magnetic diffraction pattern is obtained for the $\Gamma_5$. The ND pattern measured at 60~K and the corresponding calculated pattern with the nuclear phase alone is shown in Fig.~\ref{F4}(a), whereas the as-measured ND pattern at 5~K and the calculated pattern with the nuclear and magnetic phases are shown in Fig.~\ref{F4}(b). A good agreement between the observed and calculated patterns is evident (the reliable factors are R$_{p}$=4.29\%, R$_{wp}$=5.31\%, R$_{exp}$= 2.07\%, $\chi^2$=6.58\%, and the R$_{Mag}$= 1.84\%]. The corresponding magnetic structure is shown in Fig.~\ref{F4}(c), which is found to be pure antiferromagnetic in nature without having any net magnetization per unit cell. The magnetic moments are lying in the $ab$ plane with moment components along the $b-$axis. The magnetic structure is A-type, where all the moments in a given $ab$ plane are aligned parallel, whereas they are antiparallel to the moments in the next plane. The temperature variation of the ordered moment of Fe ions is shown in Fig.~\ref{F4}(d) where its value is estimated to be $m=$ 3.80(3) $\mu_B$/Fe$^{3+}$ at 5~K. Notably, we found an excellent agreement in the temperature value at which the moment start increasing with the magnetic susceptibility data (discussed above in Fig.~\ref{F2_R}). The temperature-dependent lattice parameters and unit cell volume are shown in Figs.~\ref{F4}(e, f) where it is clearly evident that the unit cell experiences a slight magnetostriction on cooling through the T$_N$$\approx$50 K where an expansion of the lattice constants $a$ and $c$, consequently, an expansion of unit cell volume ($V$) is also observed. Here, we find a relatively stronger magnetostriction along the $a-$axis.

\begin{figure}[h]
\includegraphics[width=3.4in]{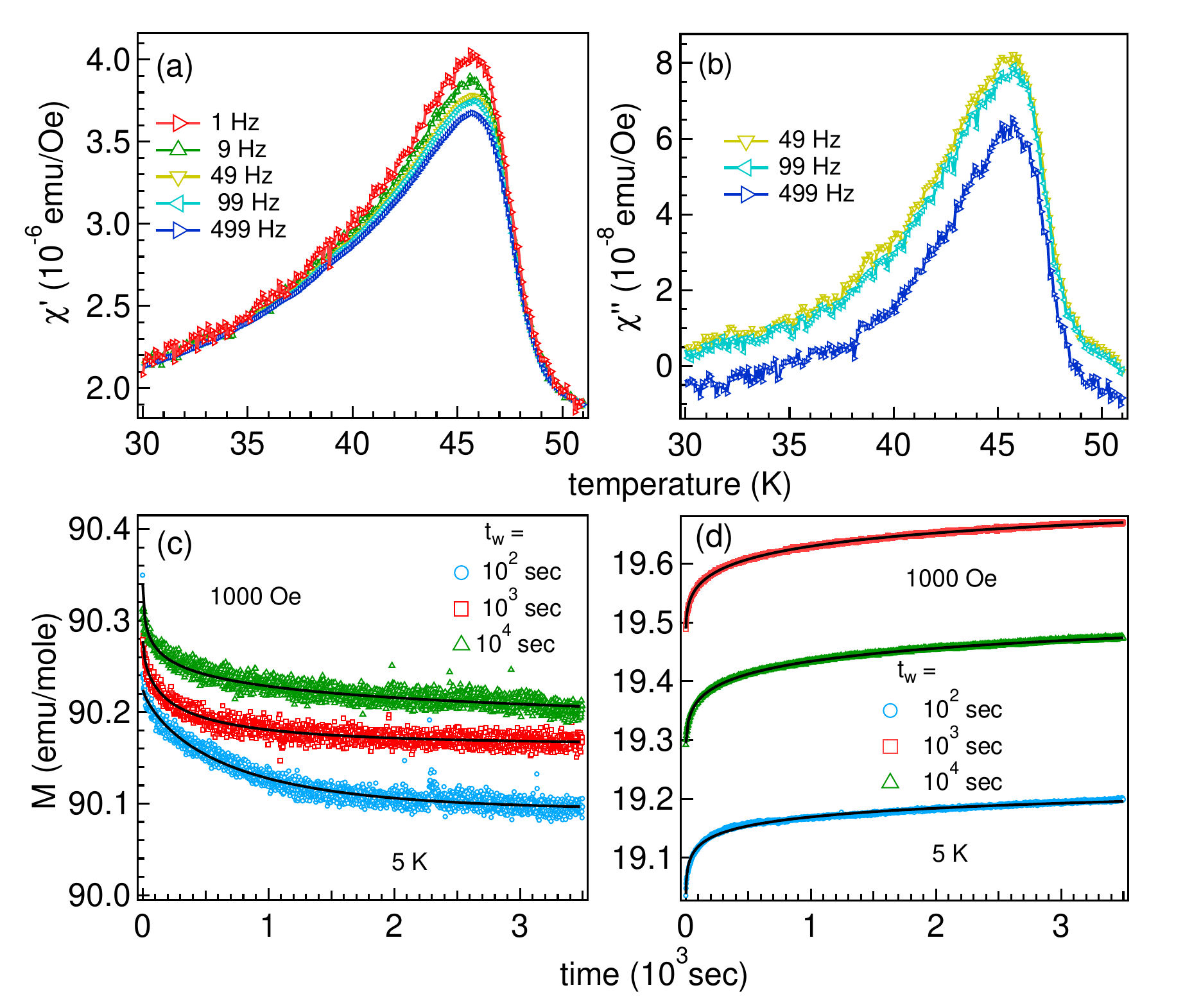} 
\caption{(a) The real and (b) imaginary part of {\it ac} magnetic susceptibility, recorded at various frequencies in the temperature range of 30--51.5~K with 2.5~Oe excitation magnetic field. (c) The FC thermoremanent magnetization (TRM) data recorded at 5~K for various waiting times (t$_w$ = 10$^2$ sec, 10$^3$ sec, and 10$^4$ sec) with a magnetic field of 1000~Oe. The solid lines depict the best data fitting using stretched exponential behavior. The semilogarithmic plot of the data for t$_w$=1000 sec is shown in the inset of (c), and the solid black line indicates the fit using the logarithmic relaxation model. (d) The ZFC time-dependent isothermal magnetization of NaFe$_2$PO$_4$(SO$_4$)$_2$ sample measured at 5 K for different waiting times (t$_w$ = 10$^2$ sec, 10$^3$ sec, and 10$^4$ sec) with a magnetic field of 1000 Oe, and the black line is depicting the best-fit curve using a stretched exponential function.} 
\label{F5}
\end{figure} 

Moreover, in Figs~\ref{F5}(a, b), the temperature dependent {\it ac} susceptibility ($\chi_{ac}= M(T)/H_{ac}$) data, measured at 2.5~Oe excitation field and at different frequencies 1, 9, 49, 99, and 499 Hz, show a clear peak centered around $\sim$46~K in both $\chi^{\prime}$--T and $\chi^{\prime\prime}$-T plots. Interestingly, a systematic reduction in the peak height is observed for both $\chi^{\prime}$ and $\chi^{\prime\prime}$ with increasing the excitation frequency, suggesting the presence of a glassy nature in the sample. However, the change in the peak position is too small to be detected with reasonable accuracy, possibly due to dominating AFM interactions as compared to the glassy nature. More importantly, the observed $\chi^{\prime\prime}$ peak, the out-of-phase component of the susceptibility, is a distinctive characteristic of glassy materials as the $\chi^{\prime\prime}$ signal is normally absent in the pure AFM systems \cite{Blundell_JPCM_19, Balanda_APP_13}. To unravel the complex spin dynamics at low temperatures, we perform the field-cooled thermo-remanent magnetization (TRM) measurement at 5~K for the different waiting times (t$_w$). In the TRM measurements, the sample was first cooled from 300~K to 5~K in the presence of 1000~Oe dc magnetic field, then after a waiting time of t$_w$, the magnetic field was turned off, and immediately started recording the magnetization as a function of time. The sample was heated up to 300~K before recording the data for another waiting time to avoid any remanence. The slow decay of the remanent magnetization with time is clearly observed in Fig.~\ref{F5}(c) and notably the magnetization increases with an increase in the waiting time. These are typical signatures of the spin glass behavior, i.e., blocking in the spins due to low temperature competing magnetic interactions in the NaFe$_2$PO$_4$(SO$_4$)$_2$. To fit the data and extract the parameters, we use the well-known stretched exponential model (with + sign) \cite{Sahoo_PRB_19}: 
\begin{equation}
M(t)=M_0 \pm M_{\mathrm{SG}} \times \exp \left[-\left(\frac{t}{t_r}\right)^{(1-\eta)}\right]
\label{TRM_E}
\end{equation}
where, M$_0$ and M$_{SG}$ are the magnetic moments of FM and spin glass components, respectively, t$_r$ is the mean relaxation time, and $\eta$ is the dispersion parameter, which depends on the relaxation rate of the spins in the glassy state. The value of the dispersion parameter is determined by the type of energy barriers encountered in the relaxation. For a system with uniform energy barriers, $\eta$=1, whereas with non-uniform energy barriers, as is usual for spin glasses, $\eta$ lies between 0 and 1. The solid black lines in Fig.~\ref{F5}(c) show the best fit using the stretched exponential model, and the fitting parameters are given in Table~S2 of \cite{supp_info}. The extracted values of t$_r$ indicate the slow relaxation of the spins, which is an evidence of the frustrated system, and the values of $\eta$ are obtained in the range of 0$<\eta<$1 for all the three different waiting times, which show the glassy magnetic phase in the sample \cite{Ajay_PRB_20_1, Markovich_PRB_10}. The glassy behavior in the NaFe$_2$PO$_4$(SO$_4$)$_2$ sample is further confirmed by the aging effect (ZFC relaxation), as shown in Fig.~\ref{F5}(d). For these measurements, the sample was first cooled down to 5~K in the zero magnetic field, then wait for a certain time (t$_w$) in the zero field. After that, a magnetic field of 1000~Oe was applied and immediately started recording the magnetization data as a function of time. The time evolution of magnetization can be nicely modeled by a stretched exponential equation~2 (with -- sign) \cite{Markovich_PRB_10}. The solid black lines in Fig.~\ref{F5}(d) represent the best fit and the fitting parameters are listed in Table~S3 of \cite{supp_info}. The values of t$_r$ and $\eta$ at 5~K are within the typical range for spin-glass systems similar to the TRM data analysis. The existence of non-zero values of both M$_0$ and M$_{\rm SG}$ confirm a mixed state of ferromagnetic and spin glass components in the NaFe$_2$PO$_4$(SO$_4$)$_2$ sample at 5~K. 

\begin{figure}[h]
\includegraphics[width=3.25in]{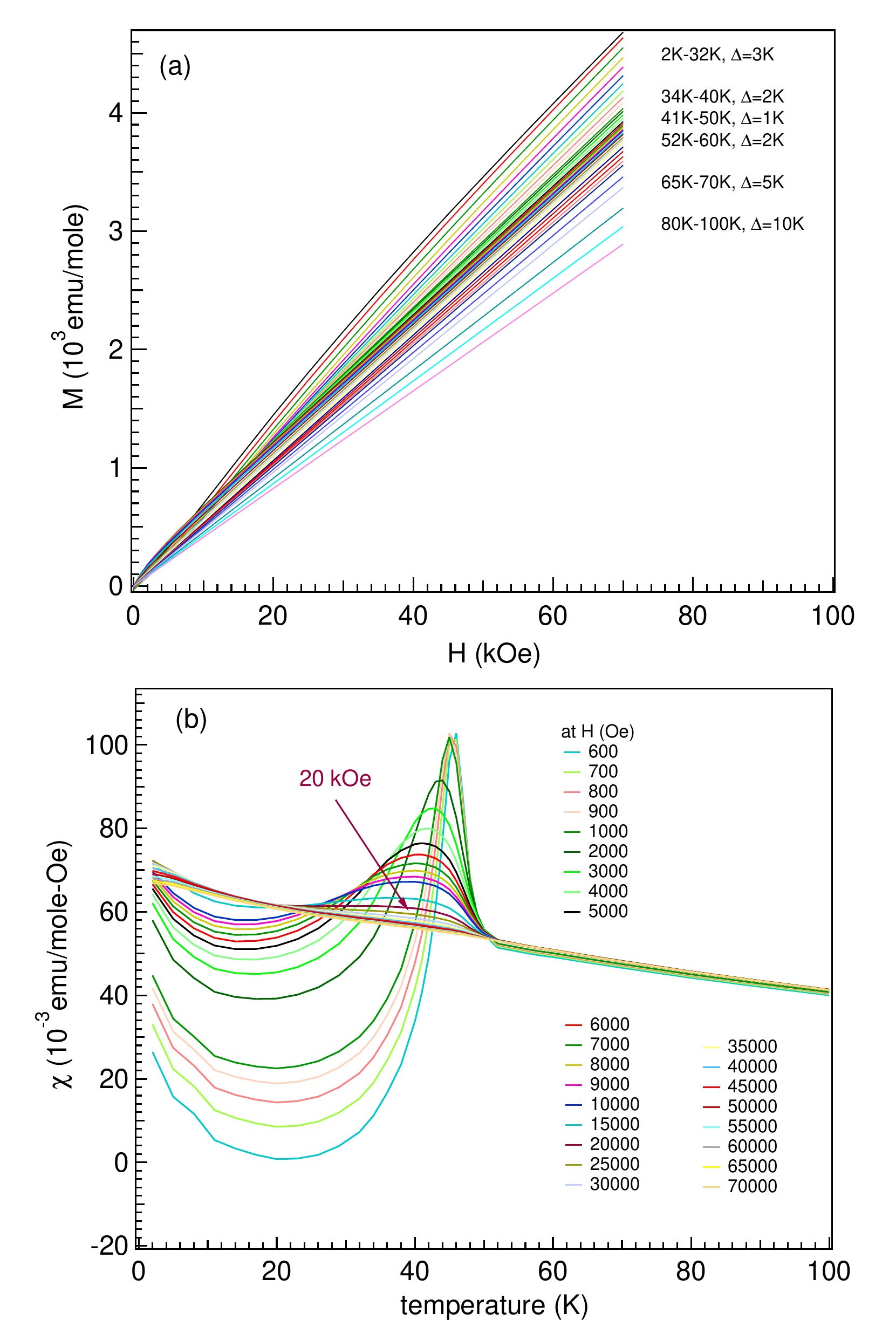}
\caption {(a) The virgin magnetization isotherms recorded from 2 to 100~K. (b) The magnetic susceptibility at different external magnetic fields extracted from the virgin isotherms.} 
\label{Fig7}
\end{figure}

\begin{figure}
\includegraphics[width=3.6in]{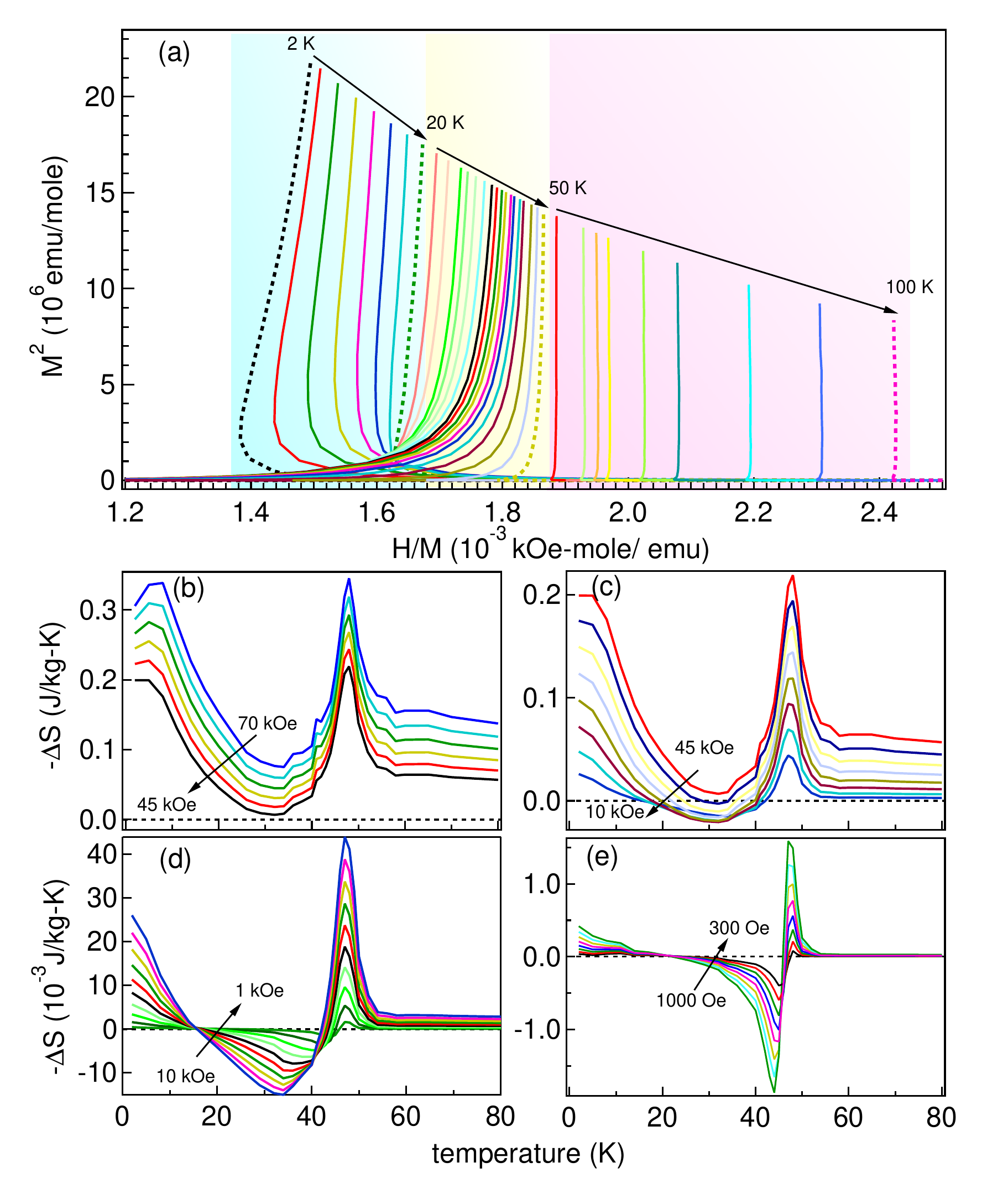}
\caption {(a) The Arrott plots extracted from the virgin data measured between 2--100~K, and the change in the magnetic entropy ($\Delta$S) versus temperature at different magnetic fields from (b) 70~kOe to 45~kOe, (c) 45~kOe to 10~kOe, (d) 10~kOe to 1~kOe, and (e) 1~kOe to 300~Oe.} 
\label{Fig8}
\end{figure}

Now it is vital to investigate the nature of magnetic correlations and magnetocaloric properties, therefore, the virgin magnetization isotherms are recorded (under ZFC mode) from 0 to 70 kOe between 2 to 100~K, as shown in Fig.~\ref{Fig7}(a). Note that in order to erase any magnetic history the sample was heated to room temperature prior to each subsequent magnetic isotherm masurement. The value of magnetization at 2~K and 70~kOe is around 0.85 $\mu_B$/f.u, which is found to be significantly lower than the theoretical value of the saturation moment of the system M$_s$= 3 $\mu_B$/Fe$^{3+}$ (g$_s$S $\mu_B$ for IS state) suggesting the AFM and/or canted spins at the low temperatures. Interestingly, we find that the isotherm curves are non-linear below the transition temperature and at lower magnetic field values, which suggest for the complex magnetic interactions in the AFM dominating glassy system \cite{Ajay_PRB_20_2}. For example, the non-linearity in these isotherms is clearly visible when we plot the $dM/dH$ versus $H$ curves, see Figs.~S2(a--d) and discussion in \cite{supp_info}. To get more insights at low temperatures, in Fig.~\ref{Fig7}(b), we plot the field-dependent susceptibility versus temperature ($\chi$--T) curves, which are retrieved from the virgin magnetization isotherms. The susceptibility curves at all fields nearly overlap in paramagnetic region above 50~K. We find that the $\chi$ values increase below 50~K and a distinctive peak emerges at $\sim$45~K, signifying the transition to AFM state \cite{Guchhait_PRB_22}. As the field strength increases, the peak height decreases and shifts towards lower temperatures along with the broadening and almost disappearing above 20~kOe. This behavior is attributed to the gradual decay of coupling at higher fields, confirming the AFM interaction/transition in the sample \cite{Yogi_PRB_17}. Also, a broad dip is observed at around 20~K for low field values, which is found to shift towards lower temperatures and then disappears above 20~kOe. At lowest measured temperature 2~K, the $\chi$ value increases with the field up to 20~kOe and then start decreasing slowly, indicating a typical behavior of AFM interactions at low temperature and low field region \cite{Vavilova_PRB_23}. Here, the magnetic moment increases at a relatively high rate with the applied field on the cost of AFM decoupling for the low field value \cite{Vavilova_PRB_23}. 

Finally, the nature of the magnetic interactions is studied with the help of Arrott plots (M$^2$ versus H/M), extracted from the isotherms across the transition \cite{Arrott_PR_57}. In Fig.~\ref{Fig8}(a), the Arrot curves exhibit vertical lines in the paramagnetic phase; however, we find positive slope in the curves below $\approx$50~K in low field region, as depicted by pink and yellow shaded areas, respectively. This indicates a second order transition from AFM to PM state. More interestingly, the slope of the Arrott plot changes to negative at temperatures below $\sim$20~K within the low magnetic field range, as displayed by the blue shaded area. To further understand the magnetic behavior of the sample in low temperatures, the change in magnetic entropy ($\Delta$S) due to the application of magnetic field is extracted from the virgin isotherms using the classical Maxwell’s thermodynamic relation, given below \cite{Magar_PRA_22, Murugan_PRB_23}
\begin{equation}
\Delta S(T, H)=\mu_0 \int_0^H\left(\frac{\partial M(T, H)}{\partial T}\right)_H d H .
\end{equation}
The $\Delta$S versus temperature (2--80~K) plot, extracted using the above equation, is shown in Figs.~\ref{Fig8}(b--d) between 300~Oe and 70~kOe. The $\Delta$S values are consistently negative for $>$45~kOe fields and exhibit a prominent peak around 48~K having the maximum value $\sim$0.34 J/kg-K at 70~kOe, see Fig.~\ref{Fig8}(b), thereby verifying the presence of the conventional magnetocaloric effect (MCE).  The $\Delta$S peaks become very sharp and undergo a cross-over from negative to positive around 48~K for 300~Oe field [see Fig.~\ref{Fig8}(e)]  revealing the occurrence of the inverse magnetocaloric effect (IMCE) \cite{Midya_PRB_84}. This cross-over temperature is found to be close to the ordering temperature obtained from magnetic susceptibility and ND analysis. Interestingly, we find that the cross-over temperature decreases with further increase in the magnetic field, as displayed in Figs.~\ref{Fig8}(c--e) due to quenching of the AFM coupling with the magnetic field. Moreover, we find another cross-over in $\Delta$S values from positive to negative at around 20~K for $<$40~kOe field values, see Figs.~\ref{Fig8}(c--e), curves  This change in the entropy curve at ~20~K seems consistent with the switching slope of the Arrott curves with temperature, as shown in Figs.~\ref{Fig8}(a),  indicating the competing magnetic interactions at low temperatures. 

\section{\noindent ~Conclusions}

	In summary, we have investigated the structural and magnetic properties of NASICON type NaFe$_2$PO$_4$(SO$_4$)$_2$ sample. The Rietveld refinement of x-ray and neutron diffraction patterns confirm the rhombohedral crystal structure, and the core-level XPS spectra and calculated effective magnetic moment confirm the presence of Fe$^{3+}$ oxidation state. The detailed analysis of temperature dependent neutron diffraction patterns reveal an AFM ordering (having zero net magnetization per unit cell) along with a magnetostriction below 50~K. Also, the unsaturated isothermal magnetization upto $\pm$70~kOe suggests for the AFM interactions; however, the observed small hysteresis in M--H loop measured at 5~K indicates the presence of weak ferromagnetic coupling at low temperatures. Interestingly, the ZFC and FC {\it d.c.}-magnetic susceptibility exhibit a distinct bifurcation at low temperatures indicating the presence of the glassy magnetic state. Also, the {\it a.c.}--magnetic susceptibility ($\chi'$ and $\chi''$) measurements, and the analysis of magnetization data at 5~K in aging and thermo-remanent modes support the existence of the glassy nature at low temperatures. The coexistence of these magnetic states may be due to the partial vacancies of Na and O ions as found in the ND and XPS studies. Interestingly, we find the complex AFM--FM interactions through the analysis of Arrott plots and magnetocaloric bahevior at low temperatures. 
	
\section{\noindent ~Acknowledgments}

M.K.S and A.K. acknowledge the MHRD and UGC, respectively, for the fellowship. We acknowledge the Department of Physics of IIT Delhi for providing XRD, Raman (DST-FIST UFO Scheme) and MPMS facilities, and central research facility (CRF) of IIT Delhi for providing XPS and PPMS facilities. This work is financially supported by SERB--DST through a core research grant (project reference no. CRG/2020/003436).

\end{document}